\documentclass[%
 apl,
 amsmath,amssymb,
preprint,
]{revtex4-1}

\usepackage{graphicx}
\usepackage{dcolumn}
\usepackage{bm}

\usepackage[utf8]{inputenc}
\usepackage[T1]{fontenc}
\usepackage{mathptmx}

\usepackage{color}
\usepackage{sidecap}

\newcommand{\YBCO}{$\mathrm{YBa}_2\mathrm{Cu}_3\mathrm{O}_7$\;}
\newcommand{\LAO}{$\mathrm{LaAl0}_3$\;}
\newcommand{\STO}{$\mathrm{SrTi0}_3$\;}

\begin{document}

\preprint{AIP/123-QED}

\title[Dynamic properties of high-$T_c$ Josephson nano-junctions made with a focused helium ion beam]{Dynamic properties of  high-$T_c$ superconducting nano-junctions made with a focused helium ion beam\\}




\author{Fran\c{c}ois Cou\"edo,$^{1}$}
\altaffiliation[Present address: ]{Laboratoire National de M\'{e}trologie et d'Essais (LNE), Quantum Electrical Metrology Department, Avenue Roger Hennequin, 78197 Trappes, France}
\author{Paul Amari,$^{1}$ Cheryl Feuillet-Palma,$^{1}$ Christian Ulysse,$^{2}$ Yogesh Kumar Srivastava,$^{3,4}$ Ranjan Singh,$^{3,4}$ Nicolas Bergeal,$^{1}$ }
\author{J\'{e}r\^{o}me Lesueur,$^{1}$}
\email[Corresponding author : ]{jerome.lesueur@espci.fr}
\affiliation{$^{1}$ Laboratoire de Physique et d'Etude des Mat\'eriaux, CNRS, ESPCI Paris, PSL Research University, UPMC, 75005 Paris, France.}
\affiliation{$^{2}$ Centre de Nanosciences et de Nanotechnologie, CNRS, Universit\'e Paris Saclay, 91120 Palaiseau, France.}
\affiliation{$^{3}$ Division of Physics and Applied Physics, School of Physical and Mathematical Sciences, Nanyang Technological University, Singapore 637371, Singapore.}
\affiliation{$^{4}$ Centre for Disruptive Photonic Technologies, The Photonics Institute, Nanyang Technological University, 50 Nanyang Avenue, Singapore 639798, Singapore.}

\date{\today}

\begin{abstract}

The Josephson junction (JJ) is the corner stone of superconducting electronics and quantum information processing. While the technology for fabricating low $T_{c}$ JJ is mature and delivers quantum circuits able to reach the "quantum supremacy", the fabrication of reproducible and low-noise high-$T_{c}$ JJ is still a challenge to be taken up. Here we report on noise properties at RF frequencies of recently introduced high-$T_{c}$ Josephson nano-junctions fabricated by mean of a Helium ion beam focused at sub-nanometer scale on a \YBCO thin film. We show that their current-voltage characteristics follow the standard Resistively-Shunted-Junction (RSJ) circuit model, and that their characteristic frequency $f_{c}=(2e/h)I_{c}R_{n}$ reaches $\sim 300$ GHz at low temperature. Using the "detector response" method, we evidence that the Josephson oscillation linewidth is only limited by the thermal noise in the RSJ model for temperature ranging from $T\sim 20$  K to $75$ K. At lower temperature and for the highest He irradiation dose, the shot noise contribution must also be taken into account when approaching the tunneling regime. We conclude that these Josephson nano-junctions present the lowest noise level possible, which makes them very promising for future applications in the microwave and terahertz regimes.

\end{abstract}

\maketitle

\section*{INTRODUCTION}
The astonishing recent evolution of Information and Communication Technologies (ICTs) is based on an accurate control of quantum properties of semiconductors at sub-micron scales. As some limitations appear, new paradigms emerge to further improve the performances of ICT devices, based on coherent quantum states and nano-scale engineering. 
Superconductivity is a very interesting platform which provides robust quantum states that can be entangled and controlled to realize quantum computation and simulation\cite{Wendin:2017bw}, or classical computation at very high speed using the so called SFQ (Single Flux Quantum) logic \cite{Tolpygo:2016hp}. This platform can also be used to make detectors of electromagnetic fields and photons operating at the quantum limit, \textit{i.e.} with unsurpassed sensitivity and resolution. These quantum sensors can be used for classical or quantum communications \cite{Holzman:2019dc}, THz waves detection and imaging \cite{Sizov:2018ii},sensitive high frequency magnetic fields measurements\cite{Clarke:2005tz,Mukhanov:2014ig}. Impressive results have been achieved in the recent years with devices based on Low critical Temperature ($T_{c}$) Superconductors (LTS) working at liquid helium temperature, and well below for Quantum Computing.

The main building block of this superconducting electronics is the Josephson Junction (JJ), a weak link between two superconducting reservoirs. While the technology for LTS JJ of typically $1\ \mu$m in size required for complex systems is mature\cite{Tolpygo:2016hp}, other ways are explored to downsize the JJ using Carbon Nano-Tubes\cite{Cleuziou:2006cs}, Copper nanowires\cite{Skryabina:2017ka} or \LAO/\STO interfaces\cite{Goswami:2016ka} for examples. The complexity and the cost of the needed cryogenic systems are clearly obstacles for large scale applications of such devices. High-$T_{c}$ superconductors (HTS) operating at moderate cryogenic temperature ($\leq 40$ K) appear as an interesting alternative solution, provided reliable JJ are available.

Different routes to make HTS JJ with suitable and reproducible characteristics are explored \cite{Mitchell:2010el,Divin:2002ky}. One of them relies on the extreme sensitivity of HTS materials such as \YBCO (YBCO) to disorder, which first reduces $T_{c}$ and then makes it insulating. High energy ion irradiation (HEII) have been used to introduce disorder in YBCO thin films through e-beam resist masks with apertures at the nanometric scale (20 - 40 nm), to make JJ\cite{Bergeal:2005jna,Bergeal:2007jc,Malnou:2014cp} and arrays\cite{Ouanani:2016cr,Pawlowski:2018jj,Couedo:2019fl} with interesting high frequencies properties, from microwaves to THz ones. Recently, Cybart \textit{et al.} successfully used a Focused Helium Ion Beam (He FIB) to locally disorder YBCO thin films and make JJ\cite{Cybart:2015cc}. In this technique, a $30$ keV He$^{+}$ ion beam of nominal size $0.7$ nm is scanned onto a thin film surface to induce disorder.
It has been used to engineer nanostructures in two-dimensional (2D) materials\cite{Iberi:2016ek,Stanford:2016bx,Zhou:2016cj}, magnetic ones\cite{Gusev:2016kd} or to make plasmonic nano-antennas for instance\cite{Scholder:2013bd}.
 Superconducting nano-structures and JJ have been fabricated with cuprate superconductors\cite{Cho:2015hq,Gozar:2017kx,Cho:2018hn,Cho:2018ds,Muller:2019uy}, MgB$_{2}$\cite{Kasaei:2018kv} and pnictides\cite{Kasaei:2019cs}. While mainly DC properties of HTS JJ made by this technique have been reported to date, the present work aims at exploring their dynamic behavior, by studying the Josephson oscillation linewidth in the tens of GHz frequency range. This characterization, which gives access to the intrinsic noise of the JJ,  is essential for high frequency applications of JJ such as mixers and detectors\cite{Malnou:2014cp,Pawlowski:2018jj,Couedo:2019fl} and to study unconventional superconductivity\cite{Kwon:2003if}.

Depending on the ion dose, HTS JJs made by the He FIB technique behave as Superconductor/Normal metal/Superconductor (SNS) JJs or Superconductor/Insulator/Superconductor (SIS) ones\cite{Cybart:2015cc}. M\"uller \textit{et al.}\cite{Muller:2019uy} evidenced scaling relations obeyed by the characteristic parameters $I_{c}$ (critical current) and $R_{n}$ (normal resistance), which are typical of highly disordered materials and known for a long time in HTS Grain-Boundary (GB) JJ\cite{Gross:1990co} for example. The large density of localized electronic defect states at the origin of this behavior is a source of $1/f$ low-frequency noise\cite{Marx:1997dw,Gustafsson:2011eea}, which broadens the Josephson oscillation linewidth at much higher frequency\cite{Likharev:1986wh}. To assess the potential performances of HTS Josephson devices made with He FIB, we directly measured the Josephson linewidth up to 40 GHz.

\section*{RESULTS}
\begin{figure*}[!htb]
\includegraphics[scale=0.8]{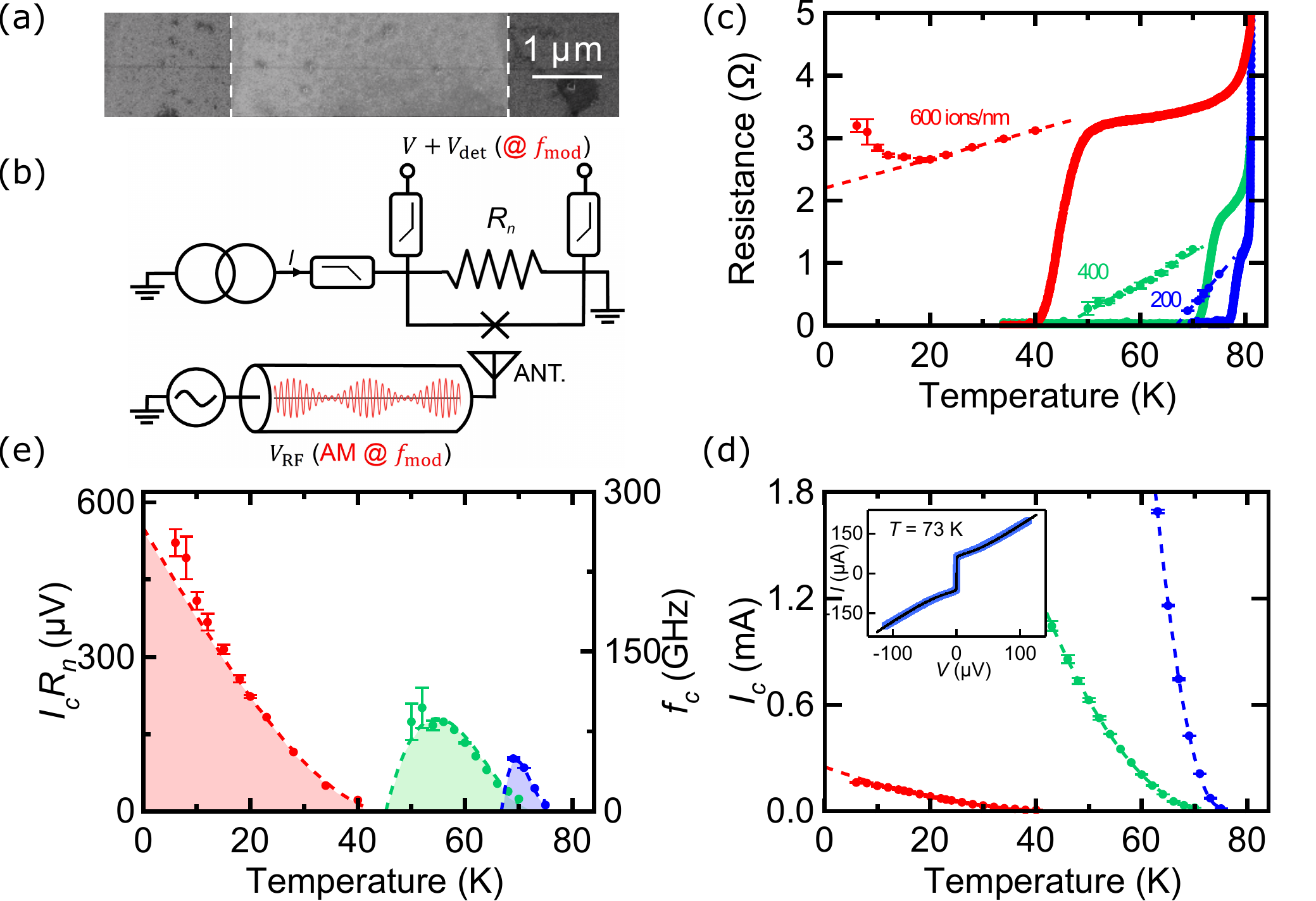}
  \caption{
  \textit{(a)} Image of a JJ using the imaging mode of the He-FIB microscope. The light grey area in between dashed lines is the superconducting channel defined by HEII The horizontal line is the 600 ions/nm dose irradiated zone, which corresponds to the barrier of the JJ.
  \textit{(b)} Sketch of the circuit used to measure the detector response signal $V_{det}$. A RF signal whose amplitude is modulated at the frequency $f_{mod}$ is sent onto the JJ via an antenna. $V_{det}$ is measured with a lock-in amplifier at $f_{mod}$. The JJ is described  according to the RSJ model, as a junction in parallel with a resistance $R_{n}$.
  \textit{(c)} $R$ vs $T$ curves (solid lines) for JJ made using different irradiation doses : $200$ ions/nm (blue), $400$ ions/nm (green) and $600$ ions/nm (red). Same color code for panels \textit{(c)} and \textit{(d)}. Below $T_{J}$, $R_{n}$ (symbols) is extracted from RSJ fits. Dashed lines show the linear decrease of $R(T)$ curves below $T_{J}$.
  \textit{(d)} $I_{c}$ vs $T$ for different irradiation doses. Dashed lines are quadratic fits. \textit{inset} : $I-V$ characteristics of a 200 ions/nm JJ. Blue line are data and black line is the RSJ fit.
  \textit{(e)} $I_{c}R_{n}$ product vs $T$ for different irradiation doses. Colored areas are calculated from the dashed lines in \textit{(c)} and \textit{(d)}, and correspond to the Josephson regimes.
  }
  \label{Figure1}
\end{figure*}

To fabricate HTS JJ, we begin with a commercial 50 nm-thick c-oriented YBCO thin film on a sapphire substrate\footnote[1]{Ceraco gmbh.} capped \textit{in-situ} with 250 nm of gold for electrical contacts. After removing the Au layer by Ar$^{+}$ ion etching everywhere except at contact pads, we structure 4 $\mu$m wide and 20 $\mu$m long channels using the HEII technique\cite{Bergeal:2005jna}. An e-beam resist mask protects the film from a 70 keV oxygen ion irradiation at a dose of $1\times10^{15}\ $ions/cm$^{2}$ to keep it superconducting. The unprotected part becomes insulating. In a second step, samples are loaded into a Zeiss Orion NanoFab Helium/Neon ion microscope and the 30 keV He$^{+}$ beam (current $\sim$ 1.15 pA) was scanned across the 4 $\mu$m-wide superconducting bridges to form JJs.
A single line is used in these experiments, whose trace can be imaged directly in the microscope \cite{Muller:2019uy} (Figure \ref{Figure1} \textit{(a)}). Imaging with the He$^{+}$ beam creates disorder, which adds to the one used to make the JJ. This is why we did not image the channels that we measured in the present study. On the same YBCO chip, we irradiated different channels with different doses ranging from 200 to 1000 ions/nm. The samples were then measured in a cryogen-free cryostat with a base temperature of 2K, equipped with filtered DC lines. The RF illumination is performed via a broadband spiral antenna placed 1 cm above the chip, and connected to a generator in a Continuous Wave (CW) mode at frequency $f$. To measure the "detector response" described below, the RF signal is electrically modulated at low frequency ($f_{mod}=199\ Hz$). 
The output signal $V_{det}$ is measured via a lock-in amplifier synchronized on this frequency.

%
Figure \ref{Figure1} \textit{(c)} shows the resistance $R$ as a function of temperature $T$ for samples irradiated with a dose of 200, 400 and 600 ions/nm. Below the $T_{c}$ of the reservoirs ($T_{c}=84\ $K), a resistance plateau develops till a transition to a zero-resistance state takes place, corresponding to the Josephson coupling through the irradiated part of the channel. Let $T_{J}$ be this coupling temperature, which decreases as the dose is increased as already reported\cite{Cybart:2015cc}\cite{Muller:2019uy}. The resistance above $T_{J}$ increases with disorder as expected, from $\lesssim 1\ \Omega$ (200 ions/nm) to $\sim 3\ \Omega$ (600 ions/nm). For irradiation dose higher than 1000 ions/nm, an insulating behavior is observed down to the lowest temperature. For the samples studied here (doses between 200 and 600 ions/nm), we measured the current-voltage ($I-V$) characteristics below $T_{J}$. The inset of Figure \ref{Figure1} \textit{(d)} shows the $I-V$ curve of the 200 ions/nm sample recorded at $T=73\ $K, which can be accurately fitted with the Resistively-Shunted-Junction (RSJ) model including thermal noise (black line), as already reported\cite{Cybart:2015cc,Muller:2019uy}. 

This model accounts for Josephson weak links and Superconductor-Normal Metal-Superconductor (SNS) junctions, where the quasiparticle current is in parallel with the superconducting one, in the limit of small junction capacitance \cite{Stewart:1968fw,Barone:1982th}. Finite temperature effect is introduced by mean of a noise current whose power spectral density corresponds to the Johnson noise of the normal state resistance $R_{n}$\cite{Likharev:1972vk,Likharev:1986wh} (see Methods section for more detail and numerical calculation).

\begin{figure*}[!t]
\includegraphics[scale=0.65]{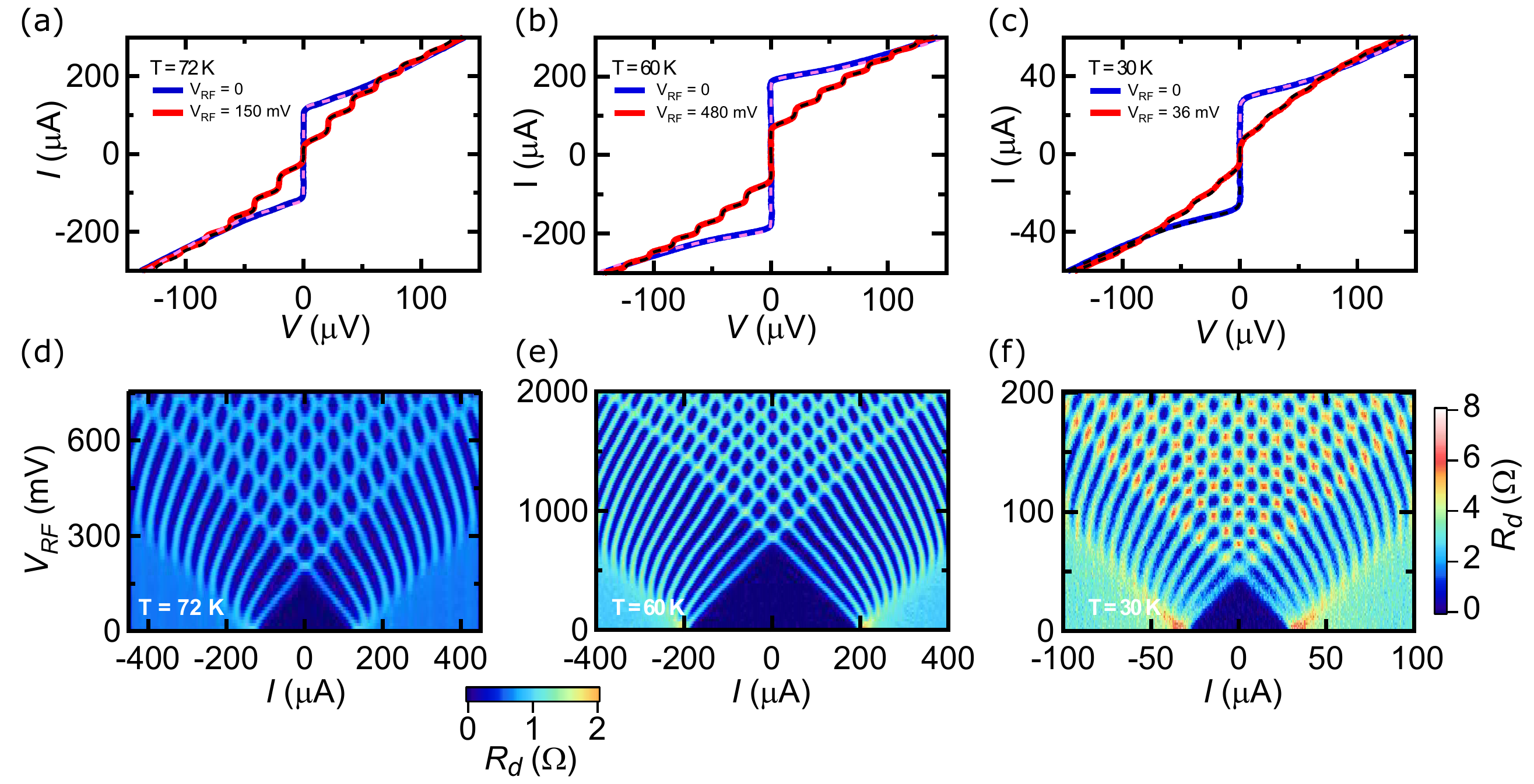}
  \caption{
  $I-V$ characteristics of \textit{(a)} a $200$ ions/nm JJ, \textit{(b)} a $400$ ions/nm JJ and \textit{(c)} a $600$ ions/nm JJ, with (red) and without (blue) $10$ GHz irradiation. Solid lines are data and dashed lines RSJ fits. 
  Color-plot of $R_{d}$ as a function of $I$ and $V_{RF}$ at $f=$10 GHz for \textit{(d)} a $200$ ions/nm JJ and \textit{(e)} a $400$ ions/nm JJ (Color-scale at the bottom).
  \textit{(f)} Same plot for a $600\ $ions/nm JJ (Color scale on the right).
  }
  \label{Figure2}
\end{figure*}

The RSJ fits are still valid when the dose and the temperature are varied, as proved by extended fits shown in Figure \ref{Figure2}. From these fits, we extracted the temperature-dependent normal-state resistance $R_{n}(T)$ and the critical current $I_{c}(T)$ presented in Figure \ref{Figure1} \textit{(c)} and \textit{(d)}, respectively, 
with an uncertainty of typically a few percents, indicated by error bars in the figures.
The former roughly follows the $R(T)$ curve measured above $T_{J}$, decreases linearly with temperature (dashed lines) and goes to zero at the superconducting temperature of the irradiated part where the Josephson regime ends. The latter has a quadratic temperature dependence (dashed lines) as expected for SNS JJ\cite{DEGENNES:1964wu, Bergeal:2005jna}. Its absolute value can exceed 1 mA (200 and 400 ions/nm doses), which corresponds to critical current densities larger than 500 kA/cm$^{2}$\cite{Muller:2019uy}. We show in Figure \ref{Figure1} \textit{(e)} the $I_{c}R_{n}$ product as a function of temperature for the different irradiation doses. 

At low doses, it shows a maximum, characteristic of SS'S junctions (where S' is a superconductor with a $T_{c}$ lower than the one of S) as observed with HEII HTS JJ\cite{Malnou:2014cp}. However, for the highest dose (600 ions/nm), it raises monotonically as the temperature is lowered. Its maximum value ($I_{c}R_{n}\sim 600\ \mu$V at 4 K) lies in between the values reported by Cybart \textit{et al.}\cite{Cybart:2015cc} and M\"{u}ller \textit{et al.}\cite{Muller:2019uy}. The corresponding characteristic frequency $f_{c}=I_{c}R_{n}/\Phi_{0}\sim 300$ GHz ($\Phi_{0}=h/2e$ the flux quantum) is higher than the one obtained by the HEII technique, which is promising for operations up to the THz frequency range\cite{Malnou:2014cp}.

We now focus on properties of these JJ at frequencies $f$ much lower than $f_{c}$, and more specifically on the Shapiro steps which appear on $I-V$ characteristics at voltages $V_{n}=n\cdot f\cdot \Phi_{0}$ ($n$ is an integer). 
Figure \ref{Figure2} \textit{(a)} shows the $I-V$ curves of the 200 ions/nm JJ measured at T=72 K without (blue) and with (red) RF irradiation at $f=10$ GHz, where we observe clear Shapiro steps. Both curves are well fitted with the RSJ model (dashed lines) with the following parameters : $R_{n}=0.5\ \Omega$, $I_{c}=133\ \mu$A and for the RF curve : $I_{RF}=133\ \mu$A (the RF current). For this temperature close to $T_{J}=75$ K, $R_{n}$ does not depend on the bias current $I$, contrary to the HEII HTS JJ\cite{Kahlmann:1998dc,Ouanani:2016cr}. Sweeping both the RF voltage $V_{RF}$ and the bias current $I$, we recorded the $I-V$ curves from which we  computed numerically the differential resistance $R_{d}=\frac{dV}{dI}$. The result is presented in color-scale in Figure \ref{Figure2} \textit{(d)}. The observation of well-defined and high-index (up to $n$=12) Shapiro steps attests the quality of this SNS JJ. Similar measurements were performed on the other JJs. The results are shown in Figure \ref{Figure2} \textit{(b)} and \textit{(c)} for the 400 ions/nm JJ and the 600 ions/nm, respectively. In each case, the measurement temperature ($T$=60 K and $T$=30 K) are close to their respective $T_{J}$. In this regime, all the curves are very well fitted with the RSJ model with the following parameters : $R_{n}=0.68\ \Omega$, $I_{c}=205\ \mu$A and $I_{RF}=143.5\ \mu$A for the 400 ions/nm JJ, and $R_{n}=2.9\ \Omega$, $I_{c}=32.5\ \mu$A and $I_{RF}=26.2\ \mu$A for the 600 ions/nm one. It is important to note that the RSJ fits were performed while taking a noise temperature equals to the bath temperature. 
Figures \ref{Figure2} \textit{(e)} and \textit{(f)} show color-plot for the corresponding samples. In both cases, pronounced oscillations with RF voltage corresponding to high order Shapiro steps are observed.

Shapiro steps unveil the internal Josephson oscillation that is produced when a JJ is biased beyond its critical current. The width of the steps is the linewidth of the Josephson oscillation\cite{Likharev:1986wh}. Within the RSJ model, Likharev and Semenov\cite{Likharev:1972vk,Likharev:1986wh} calculated the voltage power spectral density $S_{V}(f)$ and the resulting Josephson oscillation linewidth $\Delta f$ as follows : 
\begin{equation}
\Delta f=\frac{4\pi}{\Phi_{0}^2}\cdot k_{B}T\cdot \frac{R_{d}^2}{R_{n}}\cdot \left(1+\frac{I_{c}^2}{2I^2}\right)
\label{linewidth}
\end{equation}
This thermal $\Delta f$ is the minimum Josephson linewidth which can be measured, as any other noise source will increase this intrinsic linewidth.
Divin \textit{et al.} showed that $\Delta f$ can be measured experimentally from the Shapiro steps by mean of the "detector response" method\cite{Divin:1980ta,Divin:1983vu,Divin:1992ig}. The JJ is DC biased while the RF illumination is modulated at low frequency ($f_{mod}$)\cite{Sharafiev:2016fg}. The "detector response" signal $V_{det}$ is measured with a lock-in amplifier synchronized at $f_{mod}$, and plotted as a function of the DC voltage $V$ converted into a frequency $f$ through the Josephson relation $f=V/\Phi_{0}$. Centered on the Josephson frequency, \textit{i.e.} on the Shapiro step, an odd-symmetric structure appears, whose width (distance between the extrema) corresponds to $\Delta f$. To be more precise, Divin \textit{et al.}\cite{Divin:1980ta} showed that the inverse Hilbert transform of the quantity $g(V)=(8/\pi)\cdot (V_{det}/R_{d})\cdot I\cdot V$ is directly $S_{V}(f)$, a Lorentzian of width $\Delta f$ centered at the Josephson frequency $f$. This procedure, successfully used in LTS\cite{Divin:1983vu} and HTS\cite{Divin:1992ig} materials, allows to extract the Josephson linewidth accurately.

%
\begin{figure*}[!t]
\includegraphics[scale=0.9]{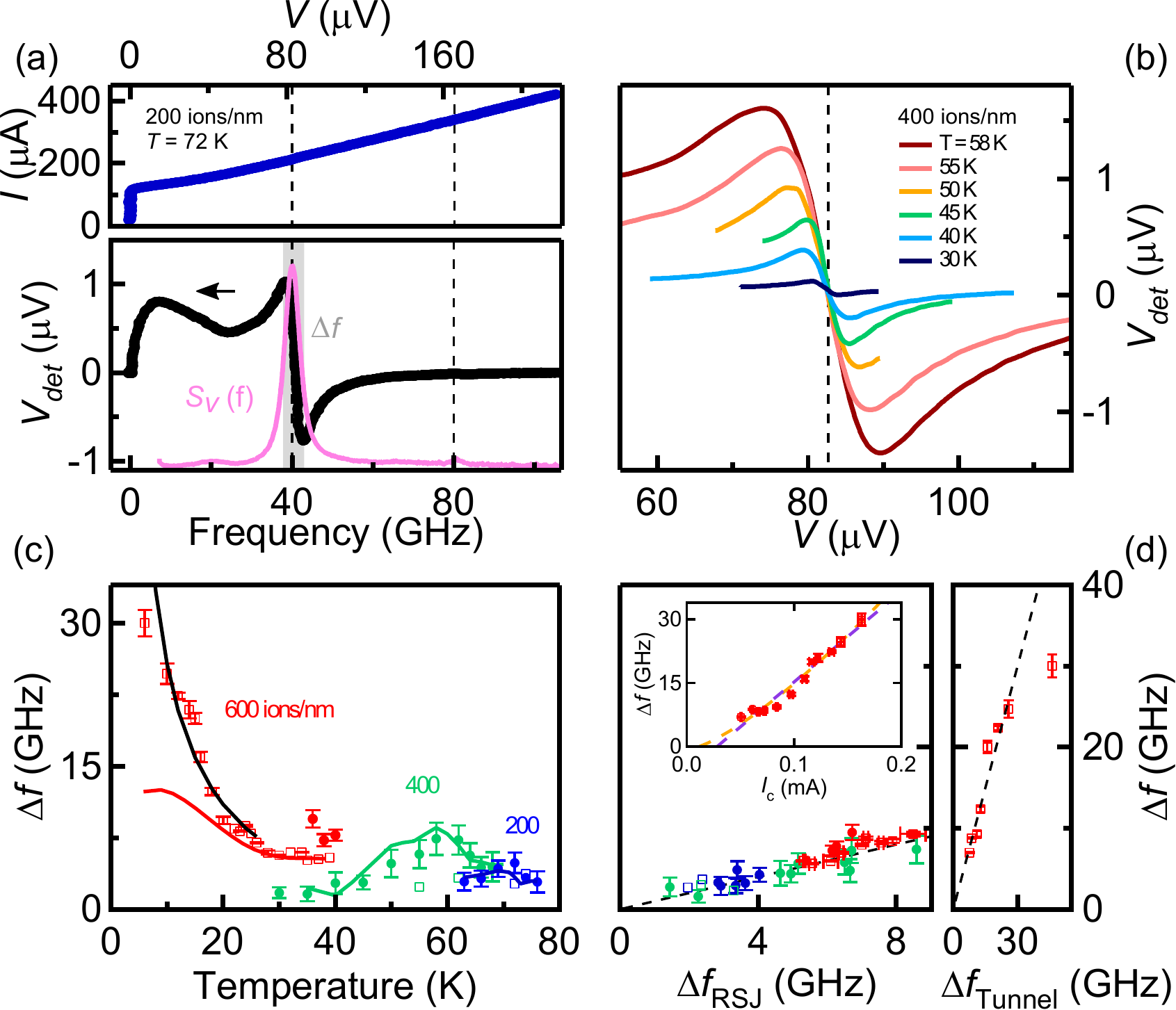}
\caption{
  \textit{(a)} $I-V$ characteristics of the $200$ ions/nm JJ under $f=40$ GHz irradiation. $V_{det}$ vs $f=V/\Phi_{0}$. The distance $\Delta f$ between the extrema corresponds to the Josephson oscillation linewidth. $S_{V}$ extracted from the inverse Hilbert transform of the the normalized response $g(V)$ (pink), whose width is $\Delta f$ (grey). 
  \textit{(b)} $V_{det}$ vs $V$ for a $400$ ions/nm JJ measured at different temperatures under $f=40$ GHz irradiation.
  \textit{(c)} $\Delta f$ vs $T$ for different irradiation doses. Solid symbols correspond to the first Shapiro step, open symbols to the second one. Solid lines (blue, green, red) are calculated from the RSJ model for $n=1$ (200 and 400 ions/nm) and $n=2$ (600 ions/nm), the black one from the tunneling one.
  \textit{(d)} $\Delta f$ vs $\Delta f_{RSJ}$ (left panel) and vs $\Delta f_{Tunnel}$ (right panel, 600 ions/nm JJ ($n=2$) for $T\leq 26 K$ ). The slope of the dashed lines is one. \textit{inset} : $\Delta f$ vs $I_{c}$ for the 600 ions/nm JJ ($n=2$) for $T\leq 26 K$. Dashed lines are best fits with a power-law exponent $1$ (purple) and $1.35$ (orange).
  }
  \label{Figure3}
\end{figure*}
We measured the Josephson oscillation linewidth of the different JJ irradiated at $f=40$ GHz using the "detector response" method. Figure \ref{Figure3} \textit{(a)} (bottom panel) shows $V_{det}$ (left axis) as a function of the frequency ($f=V/\Phi_{0}$) for the 200 ions/nm JJ. Around 40 GHz, the characteristic double-peak structure predicted by Divin\cite{Divin:1980ta} is observed, from which we extracted the Josephson linewidth ($\Delta f=4.87$ GHz). It is worth noticing that this method is highly sensitive, since the Shapiro steps cannot be seen in the $I-V$ curve simultaneously recorded (top panel). We then computed $S_{V}$ through the above explained procedure, and plotted it on the same graph (Figure \ref{Figure3} \textit{(a)} (bottom panel, right axis). Two peaks are observed, corresponding to the first an second Shapiro steps, that can be fitted with Lorentzian to extract the corresponding Josephson oscillation linewidths. For the first step (index $n=1$), the value is exactly the same as the one calculated above. Depending on the experimental conditions, we could accurately measure the Josephson linewidth of the first two Shapiro steps, or only of one of them.

We measured $V_{det}$ as a function of $V$ at different temperatures, as for example reported in Figure \ref{Figure3} \textit{(b)} for the 400 ions/nm JJ. The odd-symmetric structure at $V=82.7 \mu$V (corresponding to 40 GHz, dashed line) widens with increasing temperature as expected for thermal noise. We extracted $\Delta f$ as a function of temperature for the different samples. The result is shown in Figure \ref{Figure3} \textit{(c)}. Open (respectively solid) symbols correspond to measurements on the first (respectively second) Shapiro step. On the same graph, we added the linewidth $\Delta f_{RSJ}$ calculated for the thermal noise in the RSJ model using equation \ref{linewidth} \cite{Likharev:1972vk,Likharev:1986wh}, \textit{with no adjustable parameter}. The agreement is excellent for the 200 and 400 ions/nm JJ at all temperatures. For the 600 ions/nm, data are well reproduced at high temperature, but strongly depart from the calculation below $T=$20 K. In Figure \ref{Figure3} \textit{(d)}, we made a parametric plot of the same data : the experimental $\Delta f$ as a function of the calculated $\Delta f_{RSJ}$ in the RSJ model with thermal noise (left panel). All data align along the dashed line of slope 1, which means that noise in He FIB JJ is purely thermal, except for 600 ions/nm JJ at low temperature.

This indicates that an extra source of noise takes place below $T=$20 K in this JJ. We notice that this temperature corresponds to an up-turn in the R(T) curve (Figure \ref{Figure1} \textit{(c)}).

This thermally activated electronic transport, characteristic of a disorder-induced Anderson insulator where charge carriers hop between localized states\cite{Anderson:uc}, is well known for ion-irradiated cuprates\cite{Lesueur:1993tx}. It has been reported by Cybart \textit{et al}\cite{Cybart:2015cc} in YBCO JJ made by the He FIB technique for a sample slightly more irradiated than our 600 ions/nm one. They showed that a SIS junction is formed, and they observed a structure in the conductance related to the superconducting gap, as expected in tunnel junctions where the differential conductance is proportional to the Density of States of the reservoirs in first approximation. It is worth noting that in this regime, and contrary to the SS'S one, both $I_{c}$ and $R_{n}$ increase as the temperature is lowered, and so does the $I_{c}R_{n}$ product (see Figure \ref{Figure1} e)) to reach an interesting high value ($\sim 600\ \mu$V).
In that case, the tunneling approach proposed by Dahm \textit{et al.}\cite{Dahm:1969jm} is more appropriate than the RSJ one to calculate the Josephson oscillation linewidth, which includes the non-linear superposition of thermal and shot noises in these JJ at intermediate damping\cite{Dahm:1969jm,Likharev:1986wh}. They show that : 
\begin{equation}
\Delta f=\frac{4\pi}{\Phi_{0}^2}\cdot k_{B}T\cdot \frac{R_{d}^2}{nf\Phi_{0}}\cdot I 
\label{linewidth_SIS}
\end{equation}
We calculated $\Delta f$ for the 600 ions/nm JJ with this expression, and obtained a very good agreement with the data as shown in Figure \ref{Figure3} \textit{(c)}(black line), once again \textit{with no adjustable parameter}. The excess noise comes therefore from the shot noise contribution when approaching the tunneling limit. The parametric plot of the experimental $\Delta f$ as a function of the calculated $\Delta f_{Tunnel}$ including the shot noise (Figure \ref{Figure3} \textit{(d)} right panel) clearly shows that there is no additional noise source in our JJ.

\section*{DISCUSSION}
	This seems quite surprising, given the recent result from Müller \textit{et al.} pointing towards highly disordered JJ\cite{Muller:2019uy}, which are usually associated with strong $1/f$ Flicker noise\cite{Marx:1997dw}. They measured the low-frequency noise of SQUIDs made with low-irradiation dose JJ (230 ions/nm). A clear $1/f$ noise component is observed up to $\sim 100$ kHz. These measurements have been performed on JJ fabricated on the so-called "LSAT" substrate, which have much lower $I_{c}$ and $I_{c}R_{n}$ products than the others for the same dose. The role of the substrate on the JJ characteristics is not understood yet, but it is clear that the microstructure of the film matters for final JJ performances, and more specifically for noise properties related to defects. This has been evidenced long ago on YBCO Grain-Boundary JJ by annealing experiments\cite{Kawasaki:1992ig}. Our samples grown on sapphire may have therefore less fluctuating centers at the origin of Flicker noise than others. Past studies showed that this noise in HTS JJ is often induced by enhanced critical current fluctuations in inhomogeneous barriers\cite{Miklich:1992cx,Divin:1993do,Hao:1996kn}, and that the maximum $1/f$ noise power in the vicinity of $I_{c}$ scales with it ($S_{Vmax}\propto I_{c}^{2.7}$)\cite{Marx:1995hj}\cite{Hao:1996kn}. This  noise translates into a broader Josephson linewidth at high frequency, especially when $f<f_{c}$, \textit{i.e.} for DC bias current close to $I_{c}$\cite{Divin:1992ig}. It has been evaluated by Hao \textit{et al.}\cite{Hao:1997hm} as :
\begin{equation}
\Delta f=n\cdot \frac{2\pi}{\Phi_{0}}\cdot \sqrt{2S_{V}(f_{0})\cdot \ln{\left(\frac{f_{c}}{f_{0}}\right)}}
\label{linewidth_Hao}
\end{equation}
where $f_{0}$ is a low frequency cut-off of the $1/f$ noise (typically $f_{0}\sim 1$ Hz). Through the above mentioned scaling relation, $\Delta f$ should thus increase as $I_{c}^{1.35}$ or so. The data on our 600 ions/nm sample below $T=20$ K are compatible with this relation (inset Figure \ref{Figure3} \textit{(d)}), but we cannot make a quantitative fit since we do not know the value of $S_{V}(f_{0})$. Moreover,  the same data can be fitted with the shot noise model as well (inset Figure \ref{Figure3} \textit{(d)}), since the latter states that $\Delta f\sim I_{c}$ close to $I_{c}$ (derived from the above equation). This model fits quantitatively the evolution of $\Delta f$ with temperature with no adjustable parameter, and qualitatively the one with the critical current. We therefore conclude that the shot noise contribution fully explains the low temperature data of the most irradiated JJ, and that
there is no evidence of strong Flicker noise in the present study.

\section*{CONCLUSION}
We fabricated HTS JJ by the He FIB technique, and studied their DC and RF properties in the 10 to 40 GHz range. Their $I_{c}R_{n}$ product reaches $300$ GHz at low temperature, which is higher than for HEII JJ. We showed that Shapiro steps in the $I-V$ characteristics that appear under RF irradiation are well described by the RSJ model for SNS junctions with thermal noise. Using the "detector response" method, we determined the Josephson oscillation linewidth, and showed that it corresponds to the sole Johnson-Nyquist thermal noise in the RSJ model for the low-dose irradiated JJ. Below $T=$ 20 K, the high-dose irradiated sample has a SIS character. We demonstrated that the associated enhanced noise is due to shot noise when approaching the tunneling regime. We did not evidenced any Flicker noise component, which means that the barrier is rather homogeneous in these JJ. This study paves the way for using He FIB JJ in high frequency applications\cite{Cortez:2019bu}.


\section*{Methods}
\textbf{Resistively Shunted Junction (RSJ) model}
\subsection*{Calculation of the I-V curve at finite temperature}
The Resistively Shunted Junction (RSJ) model describes the equivalent circuit of a Josephson Junction (JJ) as two elements in parallel (the junction described by the two Josephson equations written below and its normal state resistance $R_{n}$ as sketched in Figure \ref{Figure1} (b)), biased with a current $I$\cite{Stewart:1968fw,Barone:1982th}. In this "overdamped" limit, the capacitance of the junction is neglected.
The Josephson equations state :
\begin{equation}
I_{J}=I_{c}\sin{\varphi}
\label{equationJosephson1}
\end{equation}

\begin{equation}
\frac{\partial\varphi}{\partial t}=\frac{2\pi}{\Phi_{0}}V
\label{equationJosephson2}
\end{equation}

where $I_{J}$ is the bias current and $V$ the voltage of the JJ, $I_{c}$ its critical current , $\varphi$ the quantum phase difference across it, and $\Phi_{0}$ the superconducting flux quantum.

The time evolution of the current is therefore :
\begin{equation}
I=I_{c}\sin{\varphi}+ \frac{2\pi}{\Phi_{0}R_{n}}\frac{\partial\varphi}{\partial t}
\end{equation}

The voltage is given by equation \ref{equationJosephson2}.

These equations are valid in the limit of zero-temperature. At finite temperature, the Johnson noise of the resistance must be added. The power spectral density of the current fluctuations at temperature $T$ is \cite{Likharev:1972vk,Likharev:1986wh} :
\begin{equation}
S_{I}=\frac{4k_{B}T}{R_{n}}
\label{spectraldensity}
\end{equation}

We introduce a noise current $\delta I_{n}(t)$ whose power spectral density is given by equation \ref{spectraldensity}. It has therefore a Gaussian variation in time with a variance : 
\begin{equation}
\sigma_{I}^{2}=\frac{2k_{B}T}{R_{n}\Delta t}
\label{variance}
\end{equation}

where $\Delta t$ is the time interval considered.

The time evolution of the current is now :
\begin{equation}
I+\delta I_{n}(t)=I_{c}\sin{\varphi}+ \frac{2\pi}{\Phi_{0}R_{n}}\frac{\partial\varphi}{\partial t}
\end{equation}

To get the DC I-V curve, one needs to time average this equation.

\subsection*{I-V curve in a presence of RF irradiation} 

If the JJ is submitted to a time varying current (RF irradiation) $I_{RF}=I_{RF0}\cos{(2\pi\nu_{RF}t})$ at finite temperature, the time evolution of the JJ is given by :
\begin{equation}
V(t)=R_{n}[I-I_{c}\sin{\varphi}+ \delta I_{n}(t)+I_{RF0}\cos{(2\pi\nu_{RF}t})]
\label{voltage_under_RF}
\end{equation}

and equation \ref{equationJosephson2}. After time averaging, the I-V curves present current (Shapiro) steps at voltages $V_{n}=n\Phi_{0}\nu_{RF}$, where $n$ is an integer. The width of the transition from one step to the next one is given by the thermal noise.

\subsection*{I-V curve simulation} 
In practice, the simulation of an IV curve consists in solving the equations \ref{voltage_under_RF} and \ref{equationJosephson2} by numerical integration using the Euler method. Hence the system to be numerically solved is :
\begin{equation}
V[n+1]=R_{n}(I-I_{c}\sin{\varphi[n]}+\delta I_{n}+I_{RF0}\cos{(2\pi\nu_{RF}\tau[n]}))
\end{equation}
\begin{equation}
\varphi[n+1]=\varphi[n]+\frac{2\pi}{\Phi_{0}}V[n+1]\delta\tau
\end{equation}
where the bracket notation means discrete time steps of pace $\delta\tau$, and $n$ is the step index. For each current bias $I$, a voltage vector is thus found by iteration, for each step $\delta\tau$ from $\tau=0$ to $\tau=\tau_{Max}$, starting with a random initial phase and $V[0]= 0$. The Gaussian noise $\delta I_{n}$ is a random variable changed at every step, whose variance is given by equation \ref{variance} ($\Delta t=\delta\tau$).

The system is numerically heavy to solve: first because one needs a sufficiently small $\delta\tau$ to account for the rapid variation of the voltage oscillations, especially at low bias, and at the same time one needs a sufficiently high $\tau_{Max}$ in order to have enough oscillations to average. In practice, $\delta\tau$ must be much smaller than $1/\nu_{RF}$, and $\tau_{Max}$ should be sufficiently high to average enough oscillations. We typically have vectors of 200000 points, and $\delta\tau\sim 1\cdot 10^{-12}$ s. Second, because the presence of the (actually pseudo random) noise also requires to average the calculation of each $V_{DC}$ over several iterations of the same IV curve, typically 10 times.


\section*{Aknowledgements}
The authors thank Yann Legall (ICUBE laboratory, Strasbourg) for ion irradiations. This work has been supported by the QUANTUMET ANR PRCI program (ANR-16-CE24-0028-01), the T-SUN ANR ASTRID program (ANR-13-ASTR-0025-01), the SUPERTRONICS ANR PRCE program (ANR-15-CE24-0008-03), the Emergence Program from Ville de Paris, the R\'{e}gion Ile-de-France in the framework of the DIM Nano-K and Sesame programs, the Délégation Générale à l'Armement (P. A. DGA PhD grant 2016) and the National Science Foundation Singapore (NRF2016-NRF-ANR004).

\section*{Author Contributions}
F. C. designed the samples with the help of Y. K. S. and R. S., and fabricated them with the help of P. A., C. F.-P. and C. U. F. C. performed the measurements with N. B. and C. F.-P., and most of the data analysis with J. L. J. L. and F. C. wrote the initial draft of the manuscript. All the authors contributed to the ideas behind the project, and to discussions and revisions of the manuscript.

\section*{Additional information}
Correspondence and requests for materials should be addressed to J. L.

\section*{Competing financial interests}
The authors declare no competing financial interests.

\newpage 
\section*{Bibliography}
\bibliography{FIB-He-bibliography-v3}

\newpage 

\end{document}